\documentclass[12pt]{article}

\newcommand{\TITLE}{Emerging Nonvolatile Memories for Machine Learning}

\usepackage{preamble}

\begin{document}

\maketitle

\thispagestyle{empty}

\begin{abstract}
  Digital computers have been getting exponentially faster for decades, but huge challenges exist today.
  Transistor scaling, described by Moore's law, has been slowing down over the last few years, ending the era of fully predictable performance improvements.
  Furthermore, the data-centric computing demands fueled by machine learning applications are rapidly growing, and current computing systems—even with the historical rate of improvements driven by Moore's law—cannot keep up with these enormous computational demands.
  Some are turning to analogue in-memory computing as a solution, where specialised systems operating on physical principles accelerate specific tasks.
  We explore how emerging nonvolatile memories can be used to implement such systems tailored for machine learning.
  In particular, we discuss how memristive crossbar arrays can accelerate key linear algebra operations used in neural networks, what technological challenges remain, and how they can be overcome.
\end{abstract}

\clearpage

\section{Introduction}

Machine learning has achieved impressive results in many cognitive tasks over the last $15$~years.
Computer vision (e.g.\ object detection and classification), natural language processing (e.g.\ text summarisation and chatbots), and control (e.g.\ autonomous driving) are but a few areas where statistical techniques have had a huge effect.
However, modern machine learning models have massive time and energy needs, putting practical constraints on future rapid development.
Often, the large size of these models, their energy demands and the stringent requirements of the devices (e.g.\ mobile phones) means that they cannot be run locally.
This then requires offloading the computations to powerful servers, which are interacted with through the Internet.
Although the high energy costs of models running on the servers are itself a massive problem, remote inference also introduces latency issues and privacy concerns.

The problem is an acute one because machine learning has grown at an unprecedented rate in recent years.
The success of architectures ranging from convolutional neural networks to recurrent neural networks and transformers has led to a surge of useful applications, like chatbots.
However, machine learning models are also becoming increasingly complex, leading to high computational demands even on servers with targeted optimisations.
Training the most advanced model by OpenAI, GPT-4, cost more than \$$100\,000\,000$~\cite{knightopenaiceosays}, while the cost of \emph{running} this and other models through ChatGPT interface requires an estimated \$$700\,000$/day~\cite{mokchatgptcouldcost}.

The apparent solution is to create specialised hardware that could counter the fundamental inefficiencies present in current computing systems.
One of the most pressing issues is the separation of memory and computation, thus many emerging approaches aim to perform computations directly in memory.
However, for general-purpose computing, that can be incredibly difficult to achieve, thus solutions that target only specific applications are more likely to succeed.

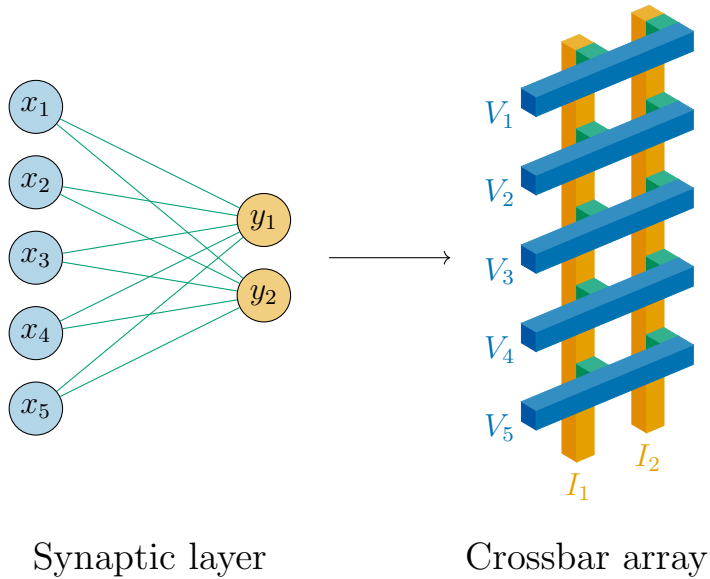
\begin{figure}[h]
    \centering
  \begin{tikzpicture}
  \tikzstyle{neuron}=[circle, draw, minimum size=20pt, inner sep=0pt]
  \tikzstyle{input neuron}=[neuron, fill=blue!30!white]
  \tikzstyle{output neuron}=[neuron, fill=orange!50!white]

  \foreach \i/\y in {1/0, 2/1, 3/2, 4/3, 5/4} {
    \node[input neuron] (input-neuron-\i) at (0, -\y) {$x_{\i}$};
  }

  \foreach \i/\y in {1/1.5, 2/2.5} {
    \node[output neuron] (output-neuron-\i) at (3, -\y) {$y_{\i}$};
  }

  \foreach \i in {1, 2, 3, 4, 5} {
    \foreach \j in {1, 2} {
      \draw[green] (input-neuron-\i) -- (output-neuron-\j);
    }
  }

  \node[right of=input-neuron-3, node distance=7cm] (crossbar-array) {
    \tdplotsetmaincoords{60}{-40}
    \begin{tikzpicture}[tdplot_main_coords]
      \def\memristorWidth{0.3}
      \def\distanceBetweenMemristors{4*\memristorWidth}
      \def\numWordLines{5}
      \def\numBitLines{2}
      \tikzmath{\wordLineLength = \numBitLines*\distanceBetweenMemristors + \memristorWidth;}
      \tikzmath{\bitLineLength = \numWordLines*\distanceBetweenMemristors + \memristorWidth;}

      \foreach \j in {1,...,\numBitLines} {
        \boxColored{\j*\distanceBetweenMemristors}{2*\memristorWidth}{0}{\memristorWidth}{\memristorWidth}{\bitLineLength}{orange}{}{black};
        \node[orange] at (\j*\distanceBetweenMemristors+0.5*\memristorWidth, 2.5*\memristorWidth, -0.5) {$I_{\j}$};
      }

      \foreach \i in {1,...,\numWordLines} {
        \foreach \j in {1,...,\numBitLines} {
          \boxColored{\j*\distanceBetweenMemristors}{\memristorWidth}{\i*\distanceBetweenMemristors}{\memristorWidth}{\memristorWidth}{\memristorWidth}{green}{}{black};
        }
      }

      \foreach \i in {1,...,\numWordLines} {
        \boxColored{0}{0}{\i*\distanceBetweenMemristors}{\wordLineLength}{\memristorWidth}{\memristorWidth}{blue}{}{black};
        \tikzmath{\myLabel = int(\numWordLines - \i + 1);}
        \node[blue] at (-0.5, 0.5*\memristorWidth, \i*\distanceBetweenMemristors+0.5*\memristorWidth) {$V_{\myLabel}$};
      }
    \end{tikzpicture}

  };

  \draw[->] ($0.5*(output-neuron-1.east) + 0.5*(output-neuron-2.east) + (0.5, 0)$) -- ($(crossbar-array.west) + (0.25, 0)$);

  \node[below of=crossbar-array, font=\large, xshift=0.25cm, yshift=-3cm] (crossbar-array-label) {Crossbar array};
  \coordinate (middle) at (1.5, 0);
  \node[font=\large] at (crossbar-array-label -| middle) {Synaptic layer};

\end{tikzpicture}%

    \caption{
      Implementing parts of a neural network using crossbar arrays.
      Software inputs $\matr{x} = \{ x_i \}$ and outputs $\matr{y} = \{ y_j \}$ are represented using physical quantities in hardware, e.g.\ voltages $\matr{V} = \{ V_i \}$ and currents $\matr{I} = \{ I_j \}$.
      The computations (like vector-matrix multiplication) are performed physically by utilising circuit laws and the structure of the crossbar array.
    }
    \label{fig:synaptic-layer-with-crossbar-array}
\end{figure}

Hardware accelerators for machine learning are a promising approach because the fundamental operations performed in structures like neural networks are simple and could potentially be implemented in memory.
Most of machine learning relies on linear algebra operations like vector-matrix products, which can be represented as series of scalar sums and multiplications.
Crossbar arrays based on emerging nonvolatile memories are a promising approach to accelerate these operations.
Using physical principles like Ohm's law and Kirchhoff's current law, relevant mathematical operations may be performed without repeatedly moving massive amounts of data.
For example, as illustrated in Figure~\ref{fig:synaptic-layer-with-crossbar-array}, synaptic layers of neural networks may be implemented using resistive crossbar arrays.

In this article, we discuss how machine learning can be accelerated using novel computing hardware based on nonvolatile memories.
We start by covering the fundamentals of conventional, digital hardware and machine learning, including today's challenges.
Then, we present the idea of analogue in-memory computing---how it can solve some of the problems of digital computer hardware, how it is implemented, and which existing technologies can help implement these analogue systems.
But because emerging memory devices based on analogue principles are not perfect, we also discuss nonidealities they suffer from, how that affects system behaviour, and what mitigation strategies can be employed.
Finally, we provide a future outlook for analogue computing.

\section{Conventional computer hardware and machine learning}

Although machine learning has proved to be a technological and commercial success, the costs associated with training and running these models are staggering.
In part, this is not surprising---machine learning systems being developed today are based on neural networks consisting of billions of parameters.
However, going forward, it is important to understand how we got here: 1)~to what extent computer hardware is responsible for the high costs, and 2)~why statistical approaches like neural networks have become so popular.

\subsection{The basics of of computer hardware}\label{sec:status-quo-of-computer-hardware}

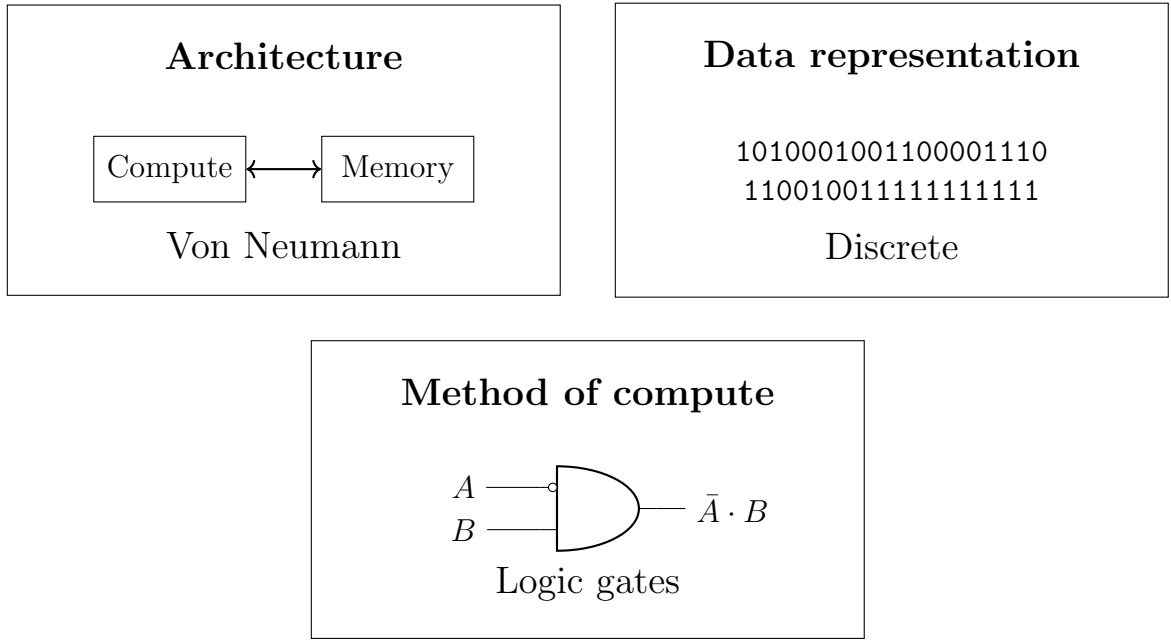
\begin{figure}[h]
    \centering
  \begin{tikzpicture}
  \tikzstyle{unit}=[draw=black, minimum width=7cm, inner ysep=0.25cm]

  \node[unit] (architecture) at (0, 0) {
    \begin{tikzpicture}
      \node[font=\large] at (0, 0) {\textbf{Architecture}};

      \node[draw=black, minimum width=2cm, minimum height=0.5cm] (compute) at (-1.5, -1.5) {Compute};
      \node[draw=black, minimum width=2cm, minimum height=0.5cm] (memory) at (1.5, -1.5) {Memory};
      \draw[<->, thick] (compute) -- (memory);

      \node[font=\large] at (0, -2.5) {Von~Neumann};
    \end{tikzpicture}
  };

  \node[unit, right of=architecture, node distance=8cm] (data-representation) {
    \begin{tikzpicture}
      \node[font=\large] at (0, 0) {\textbf{Data representation}};

      \node[align=center] at (0, -1.5) {\texttt{1010001001100001110}\\\texttt{110010011111111111}};
      \node[font=\large] at (0, -2.5) {Discrete};
    \end{tikzpicture}
  };

  \node[unit] at ($(architecture)!0.5!(data-representation) + (0, -4.5)$) (method-of-compute) {
    \begin{tikzpicture}
      \node[font=\large] at (0, 0) {\textbf{Method of compute}};

      \begin{scope}[minimum width=0cm, xshift=0.75cm, yshift=-1.5cm]
        \draw (0, 0) node[and port] (and) {};
        \node at (and.bin 1) [ocirc, left] {};
        \node (A) at ($(and.in 1) + (-1.0, 0)$) {$A$};
        \node (B) at ($(and.in 2) + (-1.0, 0)$) {$B$};
        \node (out) at ($(and.out) + (1.0, 0)$) {$\bar{A} \cdot B$};
        \draw (A) -- (and.in 1);
        \draw (B) -- (and.in 2);
        \draw (and.out) -- (out);
      \end{scope}

      \node[font=\large] at (0, -2.5) {Logic gates};
    \end{tikzpicture}
  };
\end{tikzpicture}%

    \caption{
      The operation of conventional modern computers.
    }
    \label{fig:modern-computers}
\end{figure}

For our discussion of computer hardware, three things are crucial: 1)~architecture, 2)~data representation, and 3)~method of compute.
As an example, in modern digital computers, 1)~the von~Neumann architecture is used, 2)~data are represented in a discrete way, and 3)~the computations are performed using logic gates.
This is illustrated in Figure~\ref{fig:modern-computers}.

The emergence of traditional computer architecture dates back to the $19\textsuperscript{th}$ century, when ideas about how mathematical computing devices could work were first conceived.
Charles Babbage's \emph{Analytical Engine} was an idea for a mechanical computer with an implementation that is still used today---one where memory and computation are separated.
The main reason for this separation was complexity---even individually, memory and computation units are difficult to design, thus it made sense to separate them~\cite{bromley1982babbage} but allow communication with each other.
This kind of approach was later popularised by John von~Neumann~\cite{vonneumann1993firstdraftreport} (thus the name `von~Neumann architecture') and became the norm with the emergence of the first electronic computers in the 1940s.

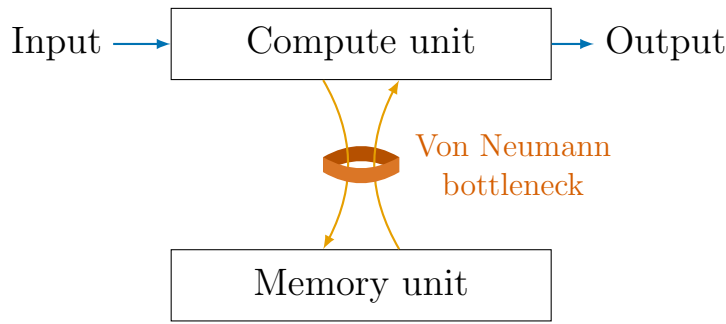
\begin{figure}[h]
    \centering
  \begin{tikzpicture}
  \tikzstyle{unit}=[draw=black, font=\large, minimum width=5cm, inner ysep=0.25cm]
  \tikzstyle{arrow}=[-latex, thick]

  \draw[fill=vermilion!85!black, draw=none] (-0.5, 0) -- ++(0, 0.2) to[bend left] ++(1.0, 0) -- ++(0, -0.2) to[bend right] ++(-1.0, 0) -- cycle;

  \node[unit] (compute-unit) at (0, 1.7) {Compute unit};
  \node[unit] (memory-unit) at (0, -1.5) {Memory unit};

  \draw[arrow, orange] ($(compute-unit.south) + (-0.5, 0)$) to[bend left] ($(memory-unit.north) + (-0.5, 0)$);
  \draw[arrow, orange] ($(memory-unit.north) + (0.5, 0)$) to[bend left] ($(compute-unit.south) + (0.5, 0)$);

  \draw[fill=vermilion!85!white, draw=none] (-0.5, 0) -- ++(0, 0.2) to[bend right] ++(1.0, 0) -- ++(0, -0.2) to[bend left] ++(-1.0, 0) -- cycle;

  \node[align=center] at (2, 0.1) {\color{vermilion}Von~Neumann\\\color{vermilion}bottleneck};

  \node[font=\large] (input) at ($(compute-unit.west) + (-1.5, 0)$) {Input};
  \node[font=\large] (output) at ($(compute-unit.east) + (1.5, 0)$) {Output};

  \draw[arrow, blue] (input) -- ($(compute-unit.west)$);
  \draw[arrow, blue] ($(compute-unit.east)$) -- (output);

\end{tikzpicture}%

    \caption{
      Von~Neumann bottleneck.
      Neural networks and other data-intensive applications are limited by the speed of data transfer between memory and compute units.
    }
    \label{fig:von-neumann-bottleneck}
\end{figure}

Although easier to implement, this architecture can lead to the so-called von~Neumann bottleneck, illustrated in Figure~\ref{fig:von-neumann-bottleneck}.
In many applications, data must be retrieved from memory and moved to computing units for the execution of mathematical operations (like multiplication and addition), and then stored back in memory.
If the application in question is data-intensive, most of the time and energy is spent on moving data between memory and compute units, rather than on the actual computations.

Another important aspect is data representation, which---at least in modern computers---has been largely defined by the concept of the Turing machine~\cite{turingComputableNumbersApplication1937}.
Alan Turing's work laid the foundation for programming---the idea that a machine can be used to implement arbitrary algorithms.
However, this all relied on the machines operating on precisely identifiable states, which in practise meant \emph{discrete} data representation.
This approach of \emph{digital} computation also largely determined how the calculations would be performed in hardware.

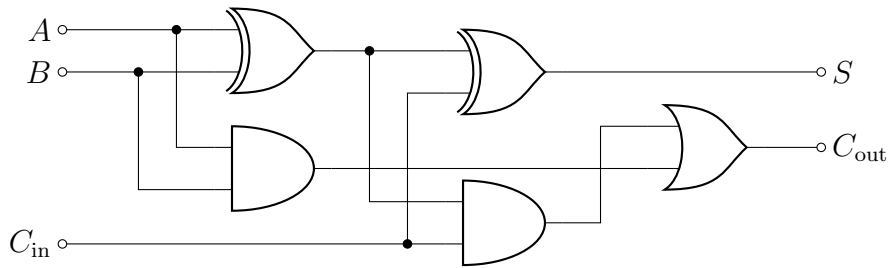
\begin{figure}[h]
  \centering
  \begin{tikzpicture}
  \node[xor port] (xor1) at (0, 0) {};

  \draw (xor1.in 1) to[short, -o] ++(-2, 0) node[left] {$A$} coordinate (A);
  \draw (xor1.in 2) to[short, -o] ++(-2, 0) node[left] {$B$} coordinate (B);

  \draw (xor1.out) to[short] ++(1.5, 0) node[xor port, anchor=in 1] (xor2) {};
  \node[and port, below of=xor2, node distance=2cm] (and2) {};

  \node[or port] (or) at ($0.5*(xor2.out)+0.5*(and2.out)+(2.5,0)$) {};

  \draw (or.in 2) to[short] (\tikztostart -| xor1.out) node[and port, anchor=out] (and1) {};

  \draw (and1.in 1) to[short] ++(-0.5, 0) to[short, -*] ($(A)!(\tikztostart)!(xor1.in 1)$);
  \draw (and1.in 2) to[short] ++(-1, 0) to[short, -*] ($(B)!(\tikztostart)!(xor1.in 2)$);

  \draw (and2.in 1) to[short] ++(-1, 0) to[short, -*] ($(xor1.out)!(\tikztostart)!(xor2.in 1)$);

  \draw (xor2.in 2) to[short] ++(-0.5, 0) to[short, -*] (\tikztostart |- and2.in 2) to[short, -o] ($(A)!(\tikztostart)!(B)$) node[left] {$C_\text{in}$};

  \draw (and2.in 2) to[short] ++(-0.5, 0);

  \draw (and2.out) to[short] ++(0.5, 0) to[short] (\tikztostart |- or.in 1) to[short] (or.in 1);

  \draw (or.out) to[short, -o] ++(0.75, 0) node[right] {$C_\text{out}$} coordinate (Cout);

  \draw (xor2.out) to[short, -o] (\tikztostart -| Cout) node[right] {$S$};
\end{tikzpicture}%

  \caption{
    Full adder circuit.
    $A$ and $B$ are the operands being added, while $C_\mathrm{in}$ is a bit carried over from the previous addition.
    $S$ is the sum, and $C_\mathrm{out}$ is the carry bit to be used in the next addition.
  }
  \label{fig:full-adder}
\end{figure}

Nowadays, logic gates are the basic building blocks of digital computers.
The theory of logic circuits was developed in parallel with the conceptualisation of the Turing machine---building on the work of George Boole, Claude Shannon and others designed switching circuits that could solve arbitrary problems in Boolean algebra~\cite{shannonSymbolicAnalysisRelay1938}.
The basic elements of these circuits---logic gates---would be implemented using mechanical relays, vacuum tubes, and, later, transistors.
But even simple operations can require a large number of devices.
For example, a full adder circuit, which adds just two binary numbers and a carry bit, shown in Figure~\ref{fig:full-adder}, requires 26~transistors~\cite{raghunandanDesignHighSpeedHybrid2019}.
Operating on a greater number of bits requires even more transistors, which in turn increases the energy consumption and the area of the circuit.

Compute optimisations in digital computers can be done at multiple levels.
At the lowest level, transistors can be scaled down to enable more complex circuits to be implemented in the same area.
Also, the overview in Figure~\ref{fig:modern-computers} does not paint the complete picture---even within the compute units, there are multiple levels of abstraction.
For example, different choices can be made about the operations that they must perform; this is often referred to as the instruction set architecture.
The widest instruction sets lead to the most general-purpose processors, like \glspl{CPU}; however, limiting the instruction set (leading to implementations like \glspl{GPU}) can result in superior performance in certain applications.
At a structural level, the conventional von~Neumann architecture remains hard to phase out, yet there have been efforts to unify memory and computation even within the digital paradigm.
For instance, microcontrollers possess a single chip that consolidates both computational components and memory.
Likewise, many cutting-edge AI chip designs strive to fit the maximum number of computational elements and memory units into one chip.

\subsection{The basics of machine learning}

Since the inception of first electro-mechanical computers, there were attempts to use them not only for precise mathematical calculations but for more open-ended and less-well-defined cognitive tasks, like classifying images.
Although in the history of machine learning, many approaches were explored, today, the most popular ones are based on statistical methods.
Specifically, artificial neural networks with their many variants have become the de facto standard for many machine learning tasks.

\begin{figure}[h]
    \centering
  \begin{tikzpicture}
  \tikzstyle{neuron}=[circle, draw, minimum size=17pt, inner sep=0pt]
  \foreach \i/\y in {1/0, 2/1, 3/2, 4/3, n/5} {
    \node[neuron, fill=blue!30] (input-neuron-\i) at (0, -\y) {$x_{\i}$};
  }
  \node[neuron, fill=blue!30] (input-neuron-bias) at (0, -6) {$1$};
  \node (ellipsis-1) at (0, -4) {$\rvdots$};

  \node[neuron, fill=green!30] (output-neuron) at (4, -2.5) {$y$};

  \foreach \i in {1, 2, 3, 4, n} {
    \draw (input-neuron-\i) -- node[fill=white, sloped] {$w_{\i}$} (output-neuron);
  }
  \draw (input-neuron-bias) -- node[fill=white, sloped] {$b$} (output-neuron);

  \node[right of=output-neuron, node distance=3.8cm] (math) {
    $
    \displaystyle
    = f \left( \sum_{i=1}^{n} \left( x_{i} w_{i} \right) + b \right)
    = f \left( \matr{x^\top} \matr{w} + b \right)
    $
  };
  \node[below of=math, node distance=1.75cm] (math) {
    where
    $
    \displaystyle
    f (z) = \begin{cases}
      1 & \text{if } z \geq 0 \\
      0 & \text{otherwise}
    \end{cases}
    $
  };
\end{tikzpicture}%

    \caption{The structure and operation of a perceptron.}
    \label{fig:perceptron}
\end{figure}
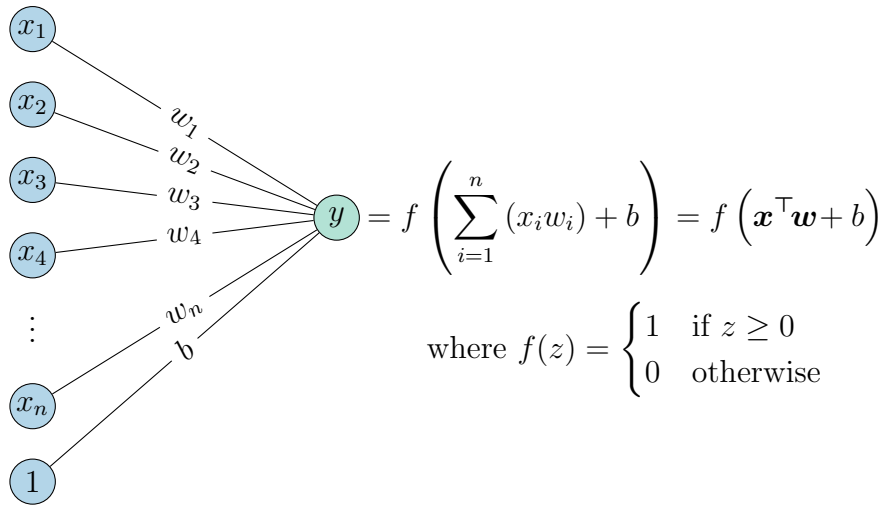

\begin{figure}[h]
    \centering
  \input{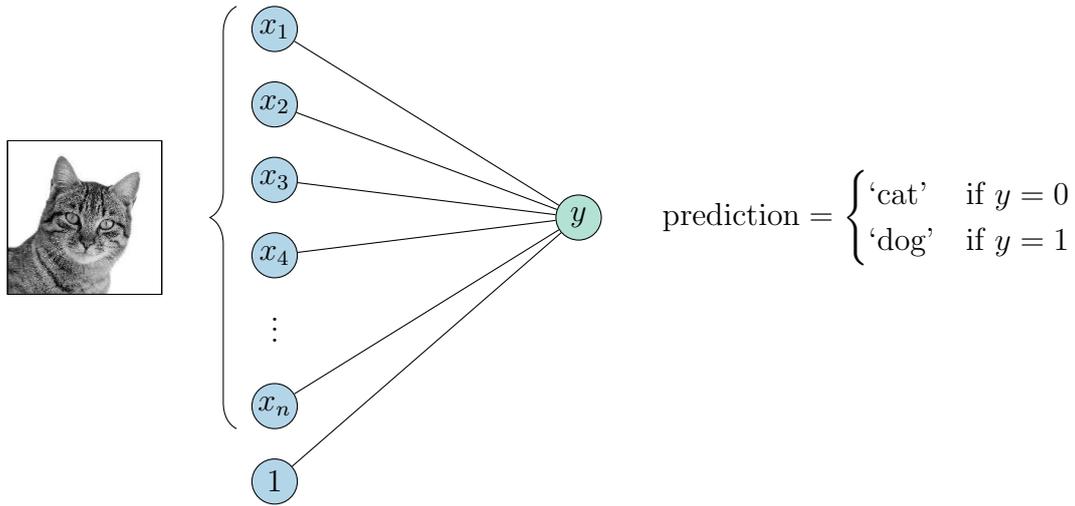}%

    \caption{Perceptron for distinguishing between dogs and cats. Pixel intensities ($0$--$255$) are used as inputs, while output $y \in \{0, 1\}$ is computed using the procedure described in Figure~\ref{fig:perceptron}.}
    \label{fig:perceptron-cat-dog}
\end{figure}

The most primitive version of artificial neural networks---the perceptron---was already conceived in the 1940s~\cite{mcculloch1943logical} and built in the 1950s~\cite{rosenblatt1957perceptron}.
As shown in Figure~\ref{fig:perceptron}, perceptrons are simple binary classifiers in which the decision is made based on a weighted sum of inputs~$\matr{x}$.
The weights~$\matr{w}$ and bias~$b$ are adjusted during training to minimise the error on the training set.
As an example, Figure~\ref{fig:perceptron-cat-dog} shows a perceptron for distinguishing between dogs and cats based on the intensity of image pixels.

For some time, however, criticisms of the perceptron~\cite{minsky1969perceptrons} led to a decline in funding and research in the broader area of connectionism.
During this period, a lot of focus was given to rules- and heuristics-based systems~\cite{lenat1983eurisko}, but such methods often lacked robustness---they had to be redesigned for individual tasks.
After advances in training algorithms, such as backpropagation~\cite{rumelhart1986learning}, and the inception of more advanced architectures, like convolutional neural networks~\cite{lecun1989handwritten}, connectionist approaches resurged in the 1990s.
Subsequent advances were in big part due to the improvement of compute devices (Moore's law) and the development of \glspl{GPU}.
The parallelisation of computations that are crucial for graphics processing (e.g.\ in computer games) proved to be useful in statistical machine learning too because the same underlying mathematical methods based on linear algebra were used.

\begin{figure}[h]
    \centering
  \begin{tikzpicture}
  \tikzstyle{neuron}=[circle, draw, minimum size=20pt, inner sep=0pt]
  \foreach \i/\y in {1/0, 2/1, 3/2, 4/3, m/5} {
    \node[neuron, fill=blue!30] (input-neuron-\i) at (0, -\y) {$x_{\i}$};
  }
  \node (ellipsis-1) at (0, -4) {$\rvdots$};
  \node[neuron, fill=blue!30] (input-neuron-bias) at (0, -6) {$1$};

  \node[neuron, fill=green!30] (output-neuron-1) at (4, -0.5) {$y_{1}$};
  \foreach \i/\y in {2/1.5, 3/2.5, n/4.5} {
    \node[neuron, fill=green!10, text=black!60, draw=black!60] (output-neuron-\i) at (4, -\y) {$y_{\i}$};
  }
  \node (ellipsis-2) at (4, -3.5) {$\rvdots$};

  \foreach \i in {1, 2, 3, 4, m, bias} {
    \foreach \j in {2, 3, n} {
      \draw[black!30] (input-neuron-\i) -- (output-neuron-\j);
    }
  }

  \foreach \i in {1, 2, 3, 4, m} {
    \draw (input-neuron-\i) -- node[fill=white, sloped, opacity=0.75, text opacity=1.0] {$w_{\i, 1}$} (output-neuron-1);
  }
  \draw (input-neuron-bias) -- node[fill=white, sloped, opacity=0.75, text opacity=1.0] {$b_1$} (output-neuron-1);

  \node[right of=output-neuron-1, node distance=4.4cm] (output) {$\displaystyle = \sigma\left(\sum_{i=1}^{m} \left( x_{i} w_{i, 1} \right) + b_1\right) = \sigma\left( \matr{x^\top} \matr{W}_{*, 1} + b_1 \right)$};

  \node[right of=output-neuron-3, node distance=7cm, yshift=-0.75cm] (output) {
    $
    \displaystyle
    \underbrace{
      \begin{bmatrix}
        y_{1} \\
        y_{2} \\
        \vdots \\
        y_{n}
      \end{bmatrix}^\top
    }_{
      \matr{y^\top}
    }
    =
    \sigma\left(
    \underbrace{
      \begin{bmatrix}
        x_{1} \\
        x_{2} \\
        \vdots \\
        x_{m}
      \end{bmatrix}^\top
    }_{
      \matr{x^\top}
    }
    \underbrace{
      \begin{bmatrix}
        w_{1, 1} & w_{1, 2} & \cdots & w_{1, n} \\
        w_{2, 1} & w_{2, 2} & \cdots & w_{2, n} \\
        \vdots & \vdots & \ddots & \vdots \\
        w_{m, 1} & w_{m, 2} & \cdots & w_{m, n}
      \end{bmatrix}
    }_{
      \matr{W}
    }
    +
    \underbrace{
      \begin{bmatrix}
        b_1 \\
        b_2 \\
        \vdots \\
        b_n
      \end{bmatrix}^\top
    }_{
      \matr{b^\top}
    }
    \right)
    $
  };

\end{tikzpicture}%

    \caption{
      Layer of a multilayer perceptron.
      $\sigma$ is a nonlinear activation function, such as the logistic function $\sigma(z) = 1 / (1 + \exp(-z))$.
      When used in a different architecture, such layers are typically called fully connected.
      Adapted from \url{https://tikz.net/neural_networks} under CC BY-SA 4.0.
    }
    \label{fig:mlp-layer}
\end{figure}
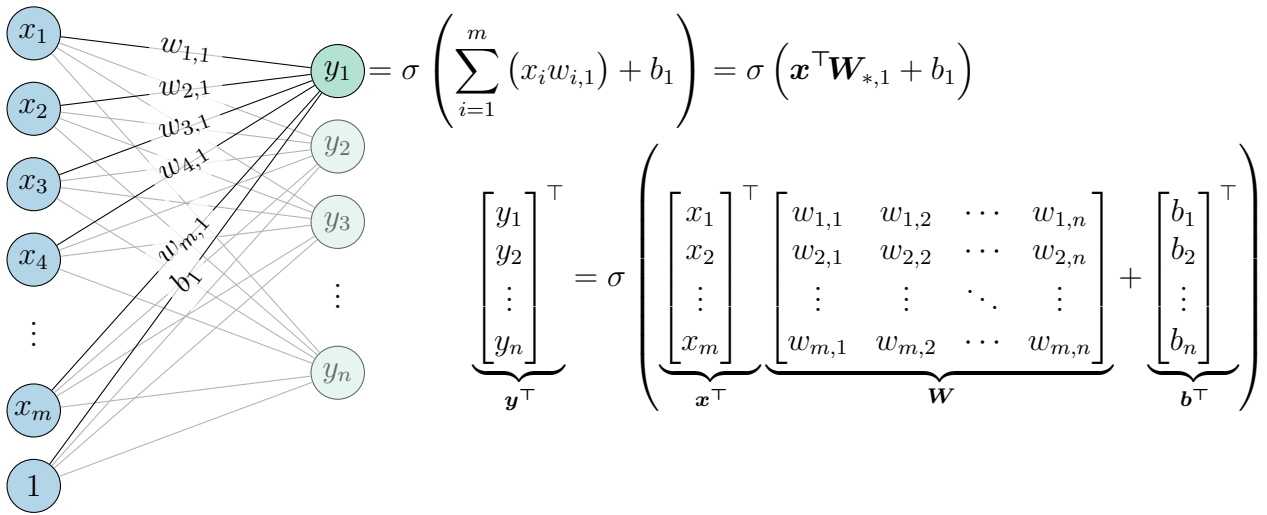

Most modern artificial neural networks have a generalisation of the perceptron as one of its parts.
These parts are called fully synaptic layers---they can have multiple postsynaptic neurons, and they also include a nonlinear activation function~$\sigma$.
Each synaptic layer consists of presynaptic neurons $\matr{x}$ and postsynaptic neurons $\matr{y}$, while the synaptic weights are stored in a matrix~$\matr{W}$---this is shown in Figure~\ref{fig:mlp-layer}.
Each postsynaptic neuron computes its activation~$y_i$ as a dot product of the presynaptic neurons and the corresponding column of the weight matrix (plus bias), followed by the activation function.
For vector $\matr{y}$ that combines all postsynaptic neurons, this can be expressed as a vector-matrix product of the inputs $\matr{x^\top}$ and weights~$\matr{W}$ (plus bias vector), followed by the activation function applied element-wise.

\begin{figure}[h]
    \centering
  \input{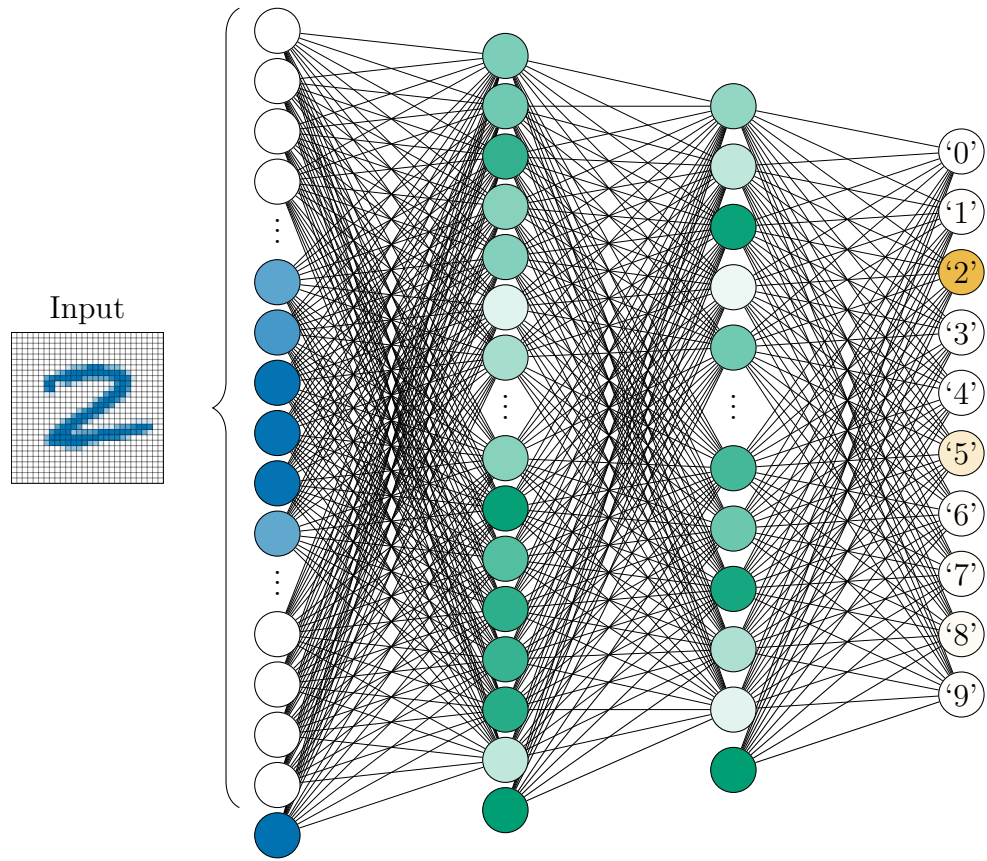}%

    \caption{%
      Multilayer perceptron.
      An input, such an image of a handwritten digit, is vectorised and fed into the first neuronal layer, the input layer, which is depicted in blue.
      The output layer (in orange) consists of as many neurons as there are classes; in this case, $10$ neurons represent the digits $0$ to $9$.
      Each neuronal layer between the input and output layers is called a hidden layer and is shown in green.
      Each neuronal layer except the output layer has a bias neuron, which is always $1$.
      }
    \label{fig:mlp}
\end{figure}

Multiple fully connected layers can be combined together to form a multilayer perceptron.
An example multilayer perceptron for recognising handwritten digits is shown in Figure~\ref{fig:mlp}.
Even if the network used is a convolutional neural network, a transformer, or some other architecture, they all tend to include these fully connected layers.

However, fully connected layers are expensive in terms of memory and computation.
For a synaptic layer with $m$ presynaptic neurons and $n$ postsynaptic neurons, the number of weights is $(m + 1) \times n$, because each presynaptic neuron is connected to each postsynaptic neuron, and there is one bias weight per postsynaptic neuron.
Although computations are expensive enough, most of the energy and time is spent on moving data between memory and compute units of the von~Neumann architecture, as discussed in Section~\ref{sec:status-quo-of-computer-hardware}---artificial neural networks are one of the classic examples of applications limited by the von~Neumann bottleneck.
The weights in the form of large matrices must be retrieved from memory every time a synaptic layer is utilised.
The problem is even more pronounced when data must be fetched from higher-capacity off-chip memory, a frequent occurrence with \emph{large} artificial neural networks.

\subsection{Status quo}

Today, computers based on the von~Neumann architecture and the \gls{CMOS} technology are ubiquitous.
Their scalability has equated to multi-decade improvements in speed and space efficiency.
However, Moore's law has been slowing down, while the costs in data-intensive applications like machine learning are starting to become prohibitive.
For example, it has been reported that programs like ChatGPT have high training~\cite{knightopenaiceosays} and inference~\cite{mokchatgptcouldcost} costs; this might be contributing to their decreased performance (as suggested by some~\cite{chenHowChatGPTBehavior2023}) in order to minimise expenses.

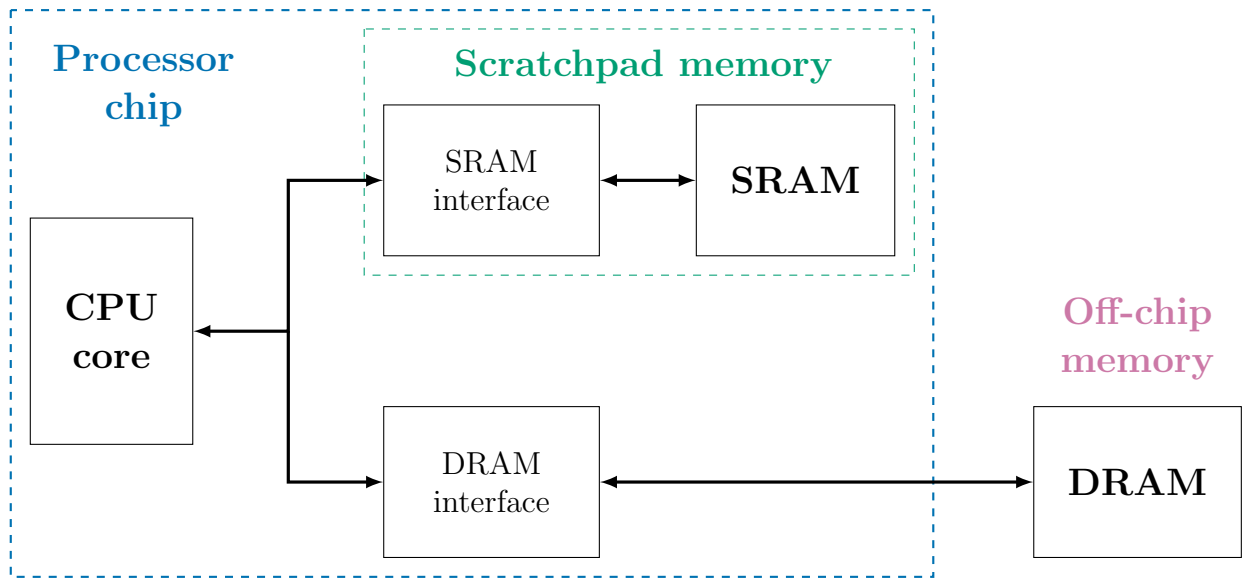
\begin{figure}[h]
    \centering
  \begin{tikzpicture}
  \tikzstyle{unit}=[draw=black, align=center, minimum height=2cm]
  \tikzstyle{main-unit}=[unit, font=\large\bfseries, inner xsep=0.45cm]
  \tikzstyle{interface-unit}=[unit, inner xsep=0.65cm]
  \tikzstyle{data-flow}=[very thick]

  \node[main-unit, minimum height=3cm] (cpu-core) at (0, 0) {CPU\\core};
  \node[interface-unit] (sram-interface) at (5, 2) {SRAM\\interface};
  \node[main-unit, right of=sram-interface, xshift=3cm] (sram) {SRAM};
  \node[interface-unit] (dram-interface) at (5, -2) {DRAM\\interface};
  \node[main-unit, right of=dram-interface, xshift=7.5cm] (dram) {DRAM};

  \draw[dashed, draw=green] ($(sram-interface.north west) + (-0.25, 1)$) rectangle ($(sram.south east) + (0.25, -0.25)$);
  \draw[thick, dashed, draw=blue] ($(cpu-core.west |- sram-interface.north) + (-0.25, 1.25)$) rectangle ($(sram.east |- dram.south) + (0.5, -0.25)$);

  \node[font=\large\bfseries, text=green] at ($0.5*(sram-interface.north) + 0.5*(sram.north) + (0, 0.5)$) {Scratchpad memory};
  \node[font=\large\bfseries, align=center, text=blue] at ($(cpu-core.west |- sram-interface.north) + (1.5, 0.25)$) {Processor\\chip};
  \node[above of=dram, yshift=0.9cm, font=\large\bfseries, align=center, text=reddish-purple] {Off-chip\\memory};

  \draw[data-flow, latex-latex] (cpu-core.east) -- ++(1.25, 0) coordinate(intersection) -- (\currentcoordinate |- sram-interface.west) -- (sram-interface.west);
  \draw[data-flow, latex-latex] (sram-interface) -- (sram);
  \draw[data-flow, -latex] (intersection) -- (\currentcoordinate |- dram-interface.west) -- (dram-interface.west);
  \draw[data-flow, latex-latex] (dram-interface) -- (dram);
\end{tikzpicture}%

    \caption{
      Typical memory architecture in modern computers.
      Compute units may utilise both off-chip memory (like \glsentryfull{DRAM}) and the more efficient on-chip memory (like \glsentryfull{SRAM}).
      The arrows denote the data flow.
      Adapted from Ref.~\cite{balasaEnergyawareMemoryManagement2016}.
    }
    \label{fig:sram-dram}
\end{figure}

Memory bottlenecks are a well-known problem, and that is reflected in modern computer architectures, a simplified example of which is shown in Figure~\ref{fig:sram-dram}.
Off-chip memory, such as \gls{DRAM}, is used for storing large amounts of data, while on-chip memory, such as \gls{SRAM}, sits closer to the compute units and is used for storing data that are frequently accessed.
On-chip memory can be more than $100$ times energy-efficient than off-chip memory~\cite{hanEIEEfficientInference2016}, but it is also more expensive and takes up more space.
That is challenging in the context of machine learning because the models being trained are incredibly large.
Without compression, even models from $10$ years ago cannot fit onto typical \gls{SRAM} units, thus the more energy-hungry \gls{DRAM} must be used~\cite{hanDeepCompressionCompressing2016}.

After memory optimisations, additional improvements can be made in the compute units.
\Glspl{CPU} are the most general-purpose computation units, which support a wide range of instructions and operate in a sequential manner.
They \emph{can} be made more parallel by adding more cores, but whether that can be successfully utilised depends on the application.
\Glspl{GPU} are more specialised, with a focus on graphics processing, but they can be used in many tasks that involve matrix operations.
They are also more parallel than \glspl{CPU} but support fewer instructions, and control flow is more limited.
\Glspl{TPU} are compute units designed specifically with machine learning in mind---optimisations are made for matrix multiplication, data retrieval, and even instruction fetching~\cite{jouppiInDatacenterPerformanceAnalysis2017}.

Finally, at the device level, improvements have tended to focus on shrinking the size of transistors, which has led to increased density and speed.
This overall trend (known as Moore's law) has led to transistors that are just a few nanometres in size.
Commercial production of \qty{5}{\nano\metre} transistors began in 2020 with the deployment in Apple, Qualcomm, Huawei, Marvell, and Nvidia consumer devices~\cite{wangInvestmentRecommendation}.
\qty{3}{\nano\metre} transistors began production in 2022 but wide-scale commercialisation has not been observed yet.
Continuous reductions in transistor size are incredibly difficult and costly to achieve, and the benefits are diminishing.
In general, Moore's law (in its traditional form) is not expected to continue simply because of physical limits---even now, the smallest transistors may be just few-tens-of-atoms-thick in certain directions.

\section{Towards \glsentrylong{IMC}}

Because further improvements at the device level in digital computers are becoming increasingly difficult, a more promising direction is alternative computing paradigms.
Specifically, targeting the aforementioned memory bottleneck could bring significant benefits.
Von~Neumann architecture results in slow and energy-inefficient data movement, which is even more pronounced today, when data-intensive applications like machine learning are becoming increasingly popular.
Bringing memory and compute units closer together---ideally achieving \gls{IMC}---could alleviate the problem.

\subsection{General overview}

Proposed solutions of addressing the von~Neumann bottleneck vary---they include digital and analogue approaches, and each can be differentiated further based on the integration level between memory and computing units.
At one end of the spectrum is the conventional von~Neumann architecture, which \emph{completely} separates memory and compute units.
Progressing towards \gls{IMC}, there exist approaches that situate memory and compute units on the same chip, e.g.\ integrating embedded \glspl{NVM} into microcontrollers, a technique often referred to as near-memory computing.
Typically, in this approach, \gls{NVM} is utilised to store model parameters, while volatile memory, like \gls{SRAM}, remains sandwiched between \gls{NVM} and compute units, holding intermediate input/output data.
In a true \gls{IMC} approach, memory serves a dual purpose: storing data and performing computations.
Achieving general-purpose computing in memory is incredibly challenging, thus a more realistic approach is to focus on specific applications.
For example, in machine learning, the goal may be to accelerate linear algebra calculations.

\begin{figure}[h]
    \centering
  \begin{tikzpicture}
  \tikzstyle{unit}=[draw=black, minimum width=7cm, inner ysep=0.25cm]

  \node[unit] (architecture-method-of-compute) at (0, 0) {
    \begin{tikzpicture}[minimum width=0cm]
      \node[font=\large, align=center] (caption) at (0, 0) {\textbf{Architecture}\\\textbf{+}\\\textbf{method of compute}};

      \node[below of=caption, node distance=3cm, inner sep=0, outer sep=0, xshift=-1.5cm] (crossbar) {
        \tdplotsetmaincoords{60}{-40}
        \begin{tikzpicture}[tdplot_main_coords]
          \def\memristorWidth{0.15}
          \def\distanceBetweenMemristors{4*\memristorWidth}
          \def\numWordLines{5}
          \def\numBitLines{2}
          \tikzmath{\wordLineLength = \numBitLines*\distanceBetweenMemristors + \memristorWidth;}
          \tikzmath{\bitLineLength = \numWordLines*\distanceBetweenMemristors + \memristorWidth;}

          \foreach \j in {1,...,\numBitLines} {
            \boxColored{\j*\distanceBetweenMemristors}{2*\memristorWidth}{0}{\memristorWidth}{\memristorWidth}{\bitLineLength}{orange}{}{black!70!white};
          }

          \foreach \i in {1,...,\numWordLines} {
            \foreach \j in {1,...,\numBitLines} {
              \boxColored{\j*\distanceBetweenMemristors}{\memristorWidth}{\i*\distanceBetweenMemristors}{\memristorWidth}{\memristorWidth}{\memristorWidth}{green}{}{black};
            }
          }

          \foreach \i in {1,...,\numWordLines} {
            \boxColored{0}{0}{\i*\distanceBetweenMemristors}{\wordLineLength}{\memristorWidth}{\memristorWidth}{blue}{}{black};
          }
        \end{tikzpicture}
      };

      \node[below of=caption, node distance=3cm, inner sep=0, outer sep=0, xshift=1.25cm] (formulas) {
        \begin{tikzpicture}
          \node (ohms-law) at (0, 0) {$\textcolor{orange}{I_{i, j}} = \textcolor{blue}{V_i} \textcolor{green}{G_{i, j}}$};
          \node[below of=ohms-law, node distance=1cm] (kirchhoffs-current-law) {$\textcolor{orange}{I_j} = \displaystyle\sum_{i} \textcolor{orange}{I_{i, j}}$};
        \end{tikzpicture}
      };

      \node[below of=formulas, font=\large, align=center, yshift=-1.5cm] (physical-laws) {Physical\\laws};
      \node[font=\large, align=center, xshift=-2.6cm] at (physical-laws) {Specialised\\circuits};

    \end{tikzpicture}
  };

  \node[unit, right of=architecture-method-of-compute, node distance=8cm] (data-representation) {
    \begin{tikzpicture}[minimum width=0cm]
      \node[font=\large] at (0, 0) {\textbf{Data representation}};

      \draw[thick] (-2.7, -1.75) -- ++(5.4, 0);

      \foreach \x in {-2.5, -2.0, ..., 2.5} {
        \draw[thick] (\x, -1.85) -- ++(0, 0.2);
      }

      \draw[thick, -latex] (-2.7, -1) -- ++(4.05, 0) -- ++(0, -0.75);

      \node[font=\large] at (0, -2.5) {Continuous};
    \end{tikzpicture}
  };
\end{tikzpicture}%

    \caption{
      The principles behind analogue compute systems.
      Unlike in Figure~\ref{fig:modern-computers}, the architecture of the system is an integral part of the computation.
    }
    \label{fig:analogue-computers}
\end{figure}
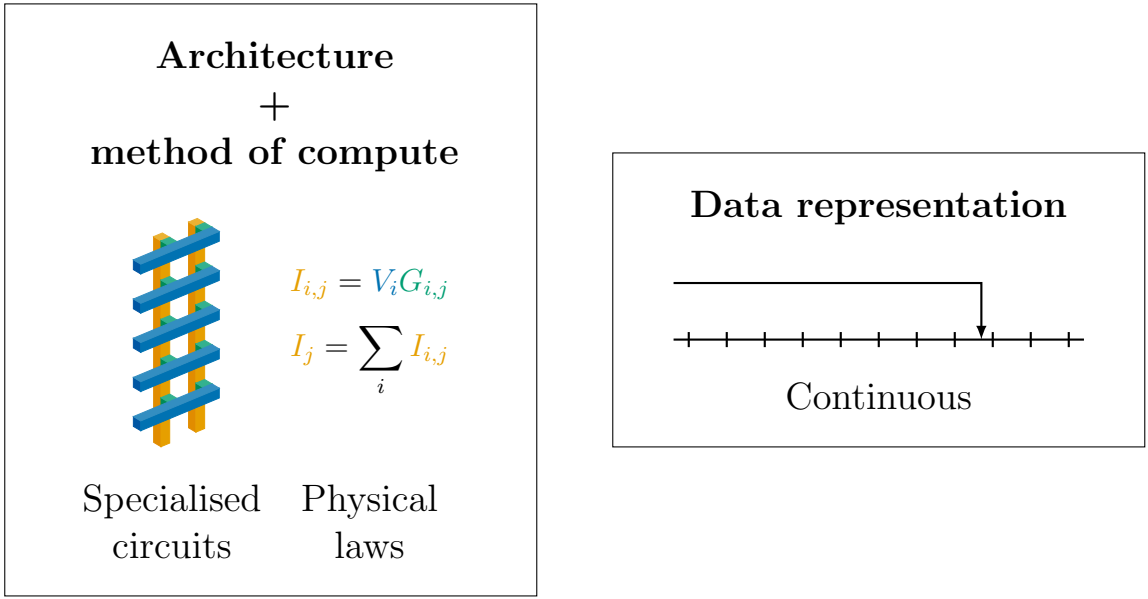

In this article, we cover the use case of accelerating the calculation of vector-matrix products, which are some of the most common operations in neural networks.
One popular \gls{IMC} approach is using resistive crossbar arrays.
Neural network weights can be represented using analogue properties, like device conductance, and the underlying physics (together with the structure of the circuit) can perform the relevant computations.
This is all done without moving the weights from memory to compute units, because the weights (in the form of conductances) are operated on directly by applying physical stimuli, like voltage, to them.
The overall approach of using analogue systems to implement \gls{IMC} and the specific example of crossbar arrays is depicted in Figure~\ref{fig:analogue-computers}.

\subsection{Implementation, from first principles}\label{sec:implementation-from-first-principles}

The idea that most analogue compute systems target specific applications can seem quite abstract.
It is thus worth exploring at least one practical example in detail to understand how this can be done.
In this section, we will focus on the aforementioned problem of computing vector-matrix products, and how it can be achieved in memory using resistive crossbar arrays.

Multiplication and addition are at the heart of linear algebra and vector-matrix products in specific---these operations can be implemented by utilising fundamental circuit laws.
For multiplication, suppose we have a resistive element with conductance $G$ (i.e.\ the reciprocal of resistance).
If we apply voltage $V$ to it, the current $I$ flowing through it will be $I = V G$, i.e.\ conductance $G$ acts as a multiplicative factor---this is known as Ohm's law.
For addition, suppose we have several branches of a circuit, each carrying current $I_i$.
When these branches meet at a junction, the total current flowing through it will be $\displaystyle I = \sum_i I_i$, i.e.\ currents are added together---this is known as Kirchhoff's current law.

\begin{figure}[h]
    \centering
  \tdplotsetmaincoords{60}{-40}
\begin{tikzpicture}[tdplot_main_coords]

  \def\memristorWidth{0.5}
  \def\distanceBetweenMemristors{4*\memristorWidth}
  \def\numWordLines{5}
  \def\numBitLines{2}
  \tikzmath{\wordLineLength = \numBitLines*\distanceBetweenMemristors + \memristorWidth;}
  \tikzmath{\bitLineLength = \numWordLines*\distanceBetweenMemristors + \memristorWidth;}

  \foreach \j in {1,...,\numBitLines} {
    \boxColored{\j*\distanceBetweenMemristors}{2*\memristorWidth}{0}{\memristorWidth}{\memristorWidth}{\bitLineLength}{orange}{}{black};

    \draw[dashed, thick, orange] (\j*\distanceBetweenMemristors, 2*\memristorWidth, -0.5*\distanceBetweenMemristors) -- ++(\memristorWidth, 0, 0) -- ++(0, \memristorWidth, 0) -- ++(-\memristorWidth, 0, 0) -- cycle;
  }

  \foreach \i in {1,...,\numWordLines} {
    \foreach \j in {1,...,\numBitLines} {
      \boxColored{\j*\distanceBetweenMemristors}{\memristorWidth}{\i*\distanceBetweenMemristors}{\memristorWidth}{\memristorWidth}{\memristorWidth}{green}{}{black};
    }
  }

  \foreach \i in {1,...,\numWordLines} {
    \boxColored{0}{0}{\i*\distanceBetweenMemristors}{\wordLineLength}{\memristorWidth}{\memristorWidth}{blue}{}{black};

    \draw[dashed, thick, blue] (-0.5*\distanceBetweenMemristors, 0, \i*\distanceBetweenMemristors) -- ++(0, 0, \memristorWidth) -- ++(0, \memristorWidth, 0) -- ++(0, 0, -\memristorWidth) -- cycle;
  }

  \tikzmath{\zV = (\numWordLines+1)*\distanceBetweenMemristors/2 + 0.5*\memristorWidth;}
  \node[rotate around z=180, canvas is yz plane at x=0, font=\LARGE] at (0.5*\distanceBetweenMemristors, -2.5*\memristorWidth, \zV) {$\color{blue}\matr{V}$};

  \tikzmath{\xI = (\numBitLines+1)*\distanceBetweenMemristors/2;}
  \node[canvas is xy plane at z=0, font=\LARGE] at (\xI, 0, -0.5*\distanceBetweenMemristors) {$\color{orange}\matr{I^\top}$};

  \tikzmath{\xG = (\numBitLines+0.55)*\distanceBetweenMemristors;}
  \tikzmath{\zG = ((\numWordLines+1)/2 + 0.25)*\distanceBetweenMemristors;}
  \node[canvas is xz plane at y=0, font=\LARGE] (G) at (\xG, 0, \zG) {$\color{green}\matr{G}$};

  \node[right of=G, node distance=3cm, font=\large] (dot-product) {
    $
    \displaystyle
    \textcolor{orange}{I_{j}} = \sum_{i} \textcolor{blue}{V_{i}} \cdot \textcolor{green}{G_{i,j}}
    $
  };

  \node[below right of=dot-product, node distance=3cm, align=center] (multiplication) {
    \large Multiplication\\
    \small Ohm's law
  };
  \draw[->, thick] (multiplication) to [in=-90,out=90] ($(dot-product) + (0.99, 0, -0.3)$);

  \node[above right of=dot-product, node distance=3.5cm, align=center, xshift=-1.5cm] (addition) {
    \small Kirchhoff's current law \\
    \large Addition
  };
  \draw[->, thick] (addition) to [in=90,out=-90] ($(dot-product) + (-0.4, 0, 0.9)$);

  \node[below of=dot-product, node distance=4cm, font=\large] (matrix-vector) {
    $
    \displaystyle
    \textcolor{orange}{\matr{I^\top}} = \textcolor{blue}{\matr{V^\top}} \textcolor{green}{\matr{G}}
    $
  };

\end{tikzpicture}%

    \caption{
      The structure and operation of a resistive crossbar array.
      Resistive elements (in green), such as memristors, are located at the intersections of word lines (in blue) and bit lines (in orange).
    When voltages $\matr{V}$ are applied to the word line, resistive element at the intersection of $i\textsuperscript{th}$ word line and $j\textsuperscript{th}$ bit line, it produces $V_i G_{i, j}$ amount of current (given zero wire resistance)---a consequence of Ohm's law.
    The currents produced by individual elements are then summed along the bit lines---a consequence of Kirchhoff's current law.
    Output current $I_j$ is thus the dot product of voltages $\matr{V}$ and the $j$-th column of the conductance matrix~$\matr{G}$.
    The vector representing \emph{all} output currents is then $\matr{I^\top} = \matr{V^\top} \matr{G}$.
    }
    \label{fig:crossbar-array}
\end{figure}
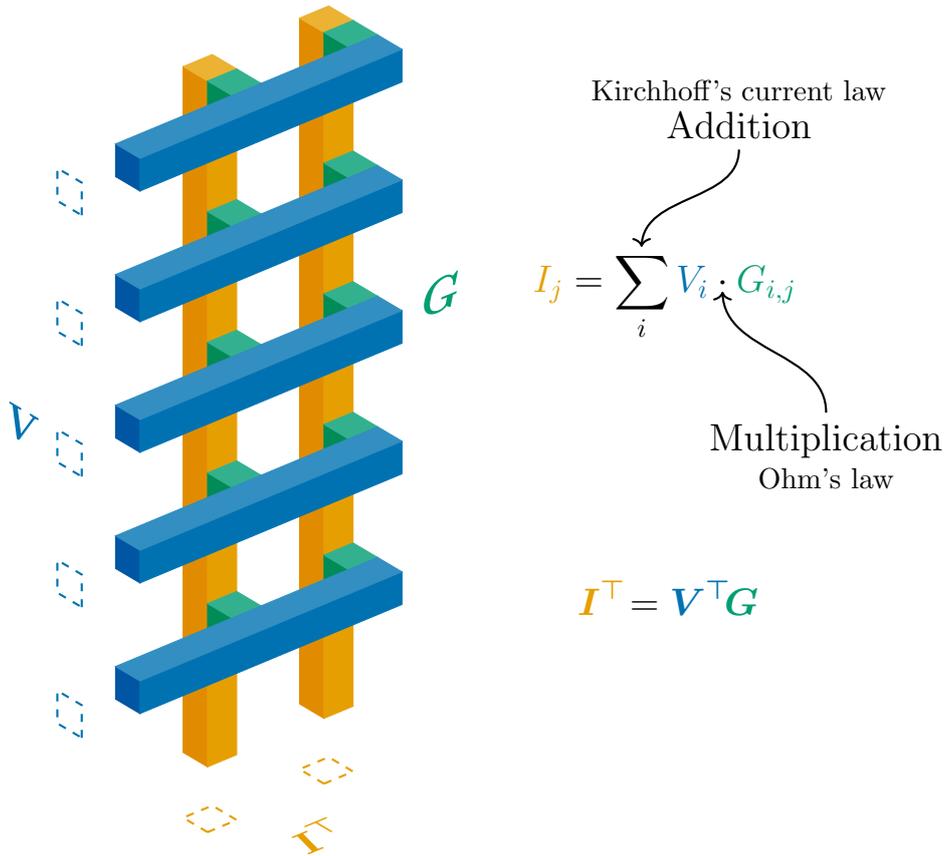

Once multiplication and addition are possible, higher-level operations can be achieved with specialised circuits.
Specifically for vector-matrix products, one can use a resistive crossbar array, which is a 2D grid of conductive wires, with resistive elements at each intersection.
Essentially, the output currents in a crossbar array are a product of a vector of voltages and a matrix of conductances, as illustrated in Figure~\ref{fig:crossbar-array}.
If we have a vector-matrix product $\matr{y} = \matr{x^\top} \matr{W}$, we can map $\matr{x}$ to voltages $\matr{V}$, $\matr{W}$ to conductances $\matr{G}$, and obtain outputs $\matr{y}$ from currents $\matr{I}$, as demonstrated in Equation~\eqref{eq:mappings}.

\begin{equation}\label{eq:mappings}
  \begin{aligned}
    \matr{V} &= k_V \matr{x} \\
    \matr{G} &= k_G \matr{W} \\
    \matr{y} &= \frac{\matr{I}}{k_V k_G}
  \end{aligned}
\end{equation}
where $k_V$ and $k_G$ are positive constants.

Crossbars can compute products of vectors of voltages and matrices of conductances because their structure determines which voltage--conductance pairs get multiplied and which resulting currents get added together.
These circuits consist of two sets of wires, called word lines and bit lines.
Voltages $\matr{V}$ are applied to the word lines, while currents $\matr{I}$ are measured on the bit lines.
A resistive element at the intersection of $i\textsuperscript{th}$ word line and $j\textsuperscript{th}$ bit line has conductance $G_{i, j}$, and when voltage $V_i$ is applied to the $i\textsuperscript{th}$ word line, the device produces $V_i G_{i, j}$ amount of current (given zero wire resistance).
The currents produced in $j\textsuperscript{th}$ bit line are summed together; the resulting current $I_j$ is the dot product of voltages $\matr{V}$ and the $j\textsuperscript{th}$ column of the conductance matrix~$\matr{G}$.
Because the $j\textsuperscript{th}$ element of a vector-matrix product is just the dot product of the vector and the $j\textsuperscript{th}$ column of the matrix, the vector representing \emph{all} output currents can be succinctly represented as $\matr{I^\top} = \matr{V^\top} \matr{G}$.

The choice of resistive devices in the crossbar array depends on the application.
In neural network \emph{training}, weights $\matr{W}$ are updated iteratively, and so the conductances in the crossbar array should have the ability to be adjusted \emph{multiple} times.
During \emph{inference}, however, the weights are fixed, and so the conductances can be fixed too after their initial programming.
Either way, the conductances will be specific to the network, so there needs to be a way to modify them \emph{at least once}.
Memristive devices (sometimes, `memristors') are typically characterised by their ability to change their conductance in response to electrical stimuli; they are thus a natural choice for crossbar-based linear algebra accelerators.
The choice of memristor may depend on whether the crossbar array is used for training or inference---the former is much more difficult and would require memristors that can be \emph{repeatedly} programmed in a \emph{linear} way.
Because of these complications, inference has been the focus of most research on memristive crossbars.

Even without considering nonidealities, any memristor will have a limited range of conductance values it can be set to, which presents problems when trying to represent real numbers $w$ using only positive conductances $G$.
Suppose the range of achievable conductances is
\begin{equation}
  G \in \interval{G_\text{off}}{G_\text{on}}
\end{equation}
Then, according to Equation~\eqref{eq:mappings}, the range of matrix values that can be represented by the crossbar array is limited too:
\begin{equation}
  w \in \interval[scaled]{\frac{G_\text{off}}{k_G}}{\frac{G_\text{on}}{k_G}}
\end{equation}
Given that $G_\text{off}$ is a positive number, only positive $w$'s can be represented.

One way to address this is using the so-called differential pairs, where matrix element $w$ is encoded into the \emph{difference} of two conductances, $G_+$ and $G_-$.
For example---and this is just \emph{one} example~\cite{JoWa2022}---the two conductances can be picked symmetrically around the `average' value~\cite{KiMa2021}:
\begin{equation}\label{eq:diff-pairs-example-mapping}
  G_\pm = G_\mathrm{avg} \pm \frac{k_G w}{2}
\end{equation}
where $G_\mathrm{avg} \defequal \frac{G_\text{off} + G_\text{on}}{2}$.

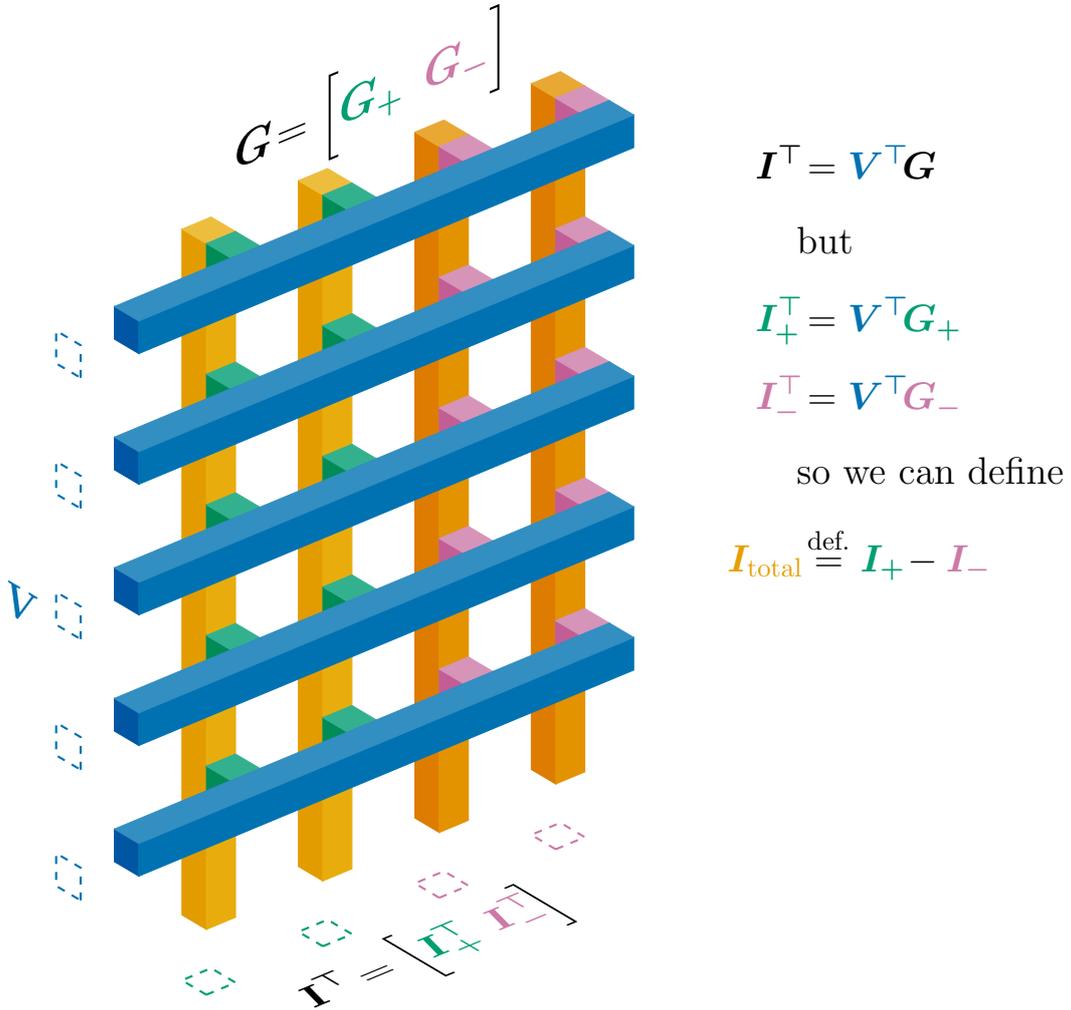
\begin{figure}[h]
    \centering
  \tdplotsetmaincoords{60}{-40}
\begin{tikzpicture}[tdplot_main_coords]

  \def\memristorWidth{0.5}
  \def\distanceBetweenMemristors{4*\memristorWidth}
  \def\numWordLines{5}
  \def\numPostsynapticNeurons{2}
  \tikzmath{\numBitLines = int(2*\numPostsynapticNeurons);}
  \tikzmath{\wordLineLength = \numBitLines*\distanceBetweenMemristors + \memristorWidth;}
  \tikzmath{\bitLineLength = \numWordLines*\distanceBetweenMemristors + \memristorWidth;}

  \def\lightColor{orange!80!yellow}
  \def\darkColor{orange!80!vermilion}

  \foreach \j in {1,...,\numPostsynapticNeurons} {
    \tikzmath{\xPos = \j*\distanceBetweenMemristors;}
    \tikzmath{\xNeg = \numPostsynapticNeurons*\distanceBetweenMemristors + \j*\distanceBetweenMemristors;}

    \boxColored{\xPos}{2*\memristorWidth}{0}{\memristorWidth}{\memristorWidth}{\bitLineLength}{\lightColor}{}{black};
    \boxColored{\xNeg}{2*\memristorWidth}{0}{\memristorWidth}{\memristorWidth}{\bitLineLength}{\darkColor}{}{black};

    \draw[dashed, thick, green] (\xPos, 2*\memristorWidth, -0.5*\distanceBetweenMemristors) -- ++(\memristorWidth, 0, 0) -- ++(0, \memristorWidth, 0) -- ++(-\memristorWidth, 0, 0) -- cycle;
    \draw[dashed, thick, reddish-purple] (\xNeg, 2*\memristorWidth, -0.5*\distanceBetweenMemristors) -- ++(\memristorWidth, 0, 0) -- ++(0, \memristorWidth, 0) -- ++(-\memristorWidth, 0, 0) -- cycle;

    \foreach \i in {1,...,\numWordLines} {
      \boxColored{\xPos}{\memristorWidth}{\i*\distanceBetweenMemristors}{\memristorWidth}{\memristorWidth}{\memristorWidth}{green}{}{black};
      \boxColored{\xNeg}{\memristorWidth}{\i*\distanceBetweenMemristors}{\memristorWidth}{\memristorWidth}{\memristorWidth}{reddish-purple}{}{black};
    }
  }

  \foreach \i in {1,...,\numWordLines} {
    \boxColored{0}{0}{\i*\distanceBetweenMemristors}{\wordLineLength}{\memristorWidth}{\memristorWidth}{blue}{}{black};

    \draw[dashed, thick, blue] (-0.5*\distanceBetweenMemristors, 0, \i*\distanceBetweenMemristors) -- ++(0, 0, \memristorWidth) -- ++(0, \memristorWidth, 0) -- ++(0, 0, -\memristorWidth) -- cycle;
  }

  \tikzmath{\zV = (\numWordLines+1)*\distanceBetweenMemristors/2;}
  \node[rotate around z=180, canvas is yz plane at x=0, font=\LARGE] at (0.5*\distanceBetweenMemristors, -2.5*\memristorWidth, \zV) {$\color{blue}\matr{V}$};

  \tikzmath{\xI = (\numBitLines+1)*\distanceBetweenMemristors/2;}
  \node[canvas is xy plane at z=0, font=\LARGE] at (\xI, 0, -0.5*\distanceBetweenMemristors) {
    $\matr{I^\top} = \begin{bmatrix} \textcolor{green}{\matr{I_+^\top}} & \textcolor{reddish-purple}{\matr{I_-^\top}} \end{bmatrix}$
    };

  \tikzmath{\xG = (\numBitLines)/2*\distanceBetweenMemristors;}
  \tikzmath{\zG = (\numWordLines+1.2)*\distanceBetweenMemristors;}
  \node[canvas is xz plane at y=0, font=\LARGE] at (\xG, 0, \zG) {$\matr{G} = \begin{bmatrix} \color{green}\matr{G_+} & \color{reddish-purple}\matr{G_-} \end{bmatrix}$};

    \tikzmath{\xEq = (\numBitLines + 2.5)*\distanceBetweenMemristors;}
    \tikzmath{\zEq = \numWordLines/2*\distanceBetweenMemristors;}
    \node[font=\large] (currents) at (\xEq, 0, \zEq) {
    $
    \displaystyle
    \begin{aligned}
      \matr{I^\top} &= \textcolor{blue}{\matr{V^\top}} \matr{G} \\[0.25cm]
      &\text{but} \\[0.25cm]
      \textcolor{green}{\matr{I_+^\top}} &= \textcolor{blue}{\matr{V^\top}} \textcolor{green}{\matr{G_+}} \\[0.25cm]
      \textcolor{reddish-purple}{\matr{I_-^\top}} &= \textcolor{blue}{\matr{V^\top}} \textcolor{reddish-purple}{\matr{G_-}} \\[0.25cm]
      &\text{so we can define} \\[0.25cm]
      \textcolor{orange}{\matr{I_\text{total}}} &\defequal \textcolor{green}{\matr{I_+}} - \textcolor{reddish-purple}{\matr{I_-}}
    \end{aligned}
    $
  };

\end{tikzpicture}%

    \caption{
      The structure and operation of a resistive crossbar array utilising differential pairs.
      Because conductances cannot be negative, one can instead encode matrix values into the difference of two separate conductance matrices $\matr{G_+}$ and $\matr{G_-}$.
      Placing them in separate bit lines of the crossbar array allows computing two separate vector-matrix products, $\matr{I_+}$ and $\matr{I_-}$.
      Assuming that an encoding scheme like the one in Equation~\eqref{eq:diff-pairs-example-mapping} has been used, subtracting $\matr{I_-}$ from $\matr{I_+}$ then yields the desired vector-matrix product $\matr{I_\text{total}}$.
      While it is not shown in the figure, this subtraction can be done directly in the circuit.
    }
    \label{fig:crossbar-array-diff-pairs}
\end{figure}

These two sets of conductances can be represented using separate conductance matrices $\matr{G_+}$ and $\matr{G_-}$, which are placed onto separate bit lines of the crossbar array, as shown in Figure~\ref{fig:crossbar-array-diff-pairs}.
These bit lines will then produce separate sets of currents, which can be represented using vectors $\matr{I_+}$ and $\matr{I_-}$.
Due to linearity of vector-matrix products, the relevant answer (from which $\matr{y}$ can be obtained by dividing by $k_V k_G$) can be computed by subtracting $\matr{I_-}$ from $\matr{I_+}$.
In practise, the `positive' and `negative' bit lines are often placed next to each other~\cite{JoFr2020}, which helps alleviate the negative effects of line resistance, a prominent nonideality (discussed in Section~\ref{sec:line-resistance}).

\subsection{Brief description of memory cells, their physical mechanisms, and performance}

It is important to understand which technologies are suitable for such \gls{IMC} systems.
When volatile memory is used for computation, the computational parameters typically need to be retrieved from nonvolatile off-chip memory, which introduces additional overhead in terms of total system performance.
The ultimate aim is to implement \gls{IMC} directly on \glspl{NVM}---this would further reduce data movement, thus enhancing energy efficiency and reducing latency.
One option is mature Flash technology, which can be utilised as computational \gls{NVM}.
However, emerging memory technologies provide advantages like compatibility with more advanced processing nodes, primarily due to lower operational voltages compared to Flash.
Additionally, many of these developing technologies offer multi-bit or even truly analogue programmability~\cite{mannocci2023inmemorycomputingemerging}.
This can be leveraged to perform multiply-accumulate operations in the analogue domain, like in Section~\ref{sec:implementation-from-first-principles}, providing further efficiency improvements. 

In a broader context, regardless of the \gls{IMC} concept, it is important to acknowledge that contemporary computing systems employ various memory technologies, generally organised in a hierarchical structure.
These memory technologies, which operate within the digital paradigm, are based on mature \gls{CMOS} fabrication and technology.
In most instances, electrical charge is used as a proxy for data, and these are called charged-based memories.
Fast volatile memory, such as \gls{SRAM}, sits at the top of this hierarchy.
It has the shortest access time, but it is also characterised by low area density and typically the highest cost.
In the middle of the hierarchy, off-chip \gls{DRAM} is usually used as the main memory.
It boasts superior area density but suffers from slightly extended access time.
At the bottom, we find \glspl{NVM} like Flash (in \glspl{SSD}) or \glspl{HDD}, which possess the highest area density but are significantly slower.
In this memory hierarchy of computing systems, emerging \gls{NVM} technologies are seen as bridging the performance gap, particularly speed, between the extremes of the hierarchy.

The leading \emph{emerging} memory technologies include \gls{RRAM}, \gls{PCM}, \gls{FeRAM}, \gls{MRAM}, \gls{STT-MRAM}, \gls{FeFET}.
However, this is not an exhaustive list; there are other promising technologies under exploration and rapid development, like \gls{ECRAM}~\cite{talin_ecram_2022}, which could offer further computational performance benefits. 

For the sake of completeness, we will briefly touch upon the physical mechanisms that underlie the functioning of these memory technologies.
Different memory cells of these emerging technologies are depicted in Figure~\ref{fig:emerging-memory-technologies}.
Regardless of the particular technology under consideration, besides the typical requirements of \gls{NVM}, the essential feature necessary for executing linear algebra operations using, for example, crossbar arrays is the capability to effectively configure a single memory cell into multiple stable, nonvolatile memory states.

\begin{figure}[h]
    \centering
  \input{figures/tikz/emerging-memory-technologies/f.tex}%

    \caption{
      Emerging memory technologies.
      (a)~\Glsentryfull{RRAM},
      (b)~\glsentryfull{PCM},
      (c)~\glsentryfull{FeRAM},
      (d)~\glsentryfull{STT-MRAM},
      (e)~\glsentryfull{FeFET},
      (f)~\glsentryfull{SOT-MRAM},
      (g)~\glsentryfull{ECRAM},
      (h)~\ce{MoS2}-channel-based memtransistor.
      Adapted from Ref.~\cite{mannocci2023inmemorycomputingemerging}.
    }
    \label{fig:emerging-memory-technologies}
\end{figure}

\Pgls{RRAM} cell is typically structured on a metal--insulator--metal basis, where the central insulating layer serves as a resistance-switching layer.
This layer is generally composed of metal oxides, but other materials like chalcogenides, organic substances, nitrides, and 2D materials are also being explored.
Additionally, a distinction can be made between filamentary and interface-based switching.
This differentiation depends on whether the write operation results in the formation of conductive filaments within the insulating environment, or if the resistance switching is based on modulating barrier heights (for example, Schottky barriers) between the switching layer and the electrodes (metals).

\Pgls{PCM} cell shares similarities with \pgls{RRAM} cell, but the switching layer in \gls{PCM} is made up of phase-change materials.
These materials have the ability to reversibly switch from crystalline to amorphous phases.
A common example of a phase-change material is chalcogenide (such as \ce{Ge2Sb2Te5}), where the phase transition is driven by Joule heating generated by voltage pulses applied across the cell.

Both \gls{RRAM} and \gls{PCM} have the capability to store intermediate states, apart from the two distinct states.
This can be achieved either by modifying the morphology of the conductive filaments (in the case of \glspl{RRAM}), or by varying the proportion of material that transitions between the two distinct phases.
This feature is of critical interest in the context of analogue or multi-bit computing.
However, several challenges persist, including the precision and practicality of programming multiple intermediate states and the retention of these states.

\Gls{FeRAM} and \gls{FeFET} memory rely on ferroelectric effects.
Typical examples of ferroelectric materials for \gls{FeRAM} are perovskites, though recently, there has been significant interest~\cite{mulaosmanovic_ferroelectric_2021} in employing ferroelectricity from \ce{HfO_x}-based materials (either doped or undoped) due to their superior \gls{CMOS} compatibility.
In the case of \gls{FeRAM}, the state is read by measuring the displacement current during switching in the ferroelectric material, which is a destructive process and not ideal for computing applications.
Alternative strategies include \glspl{FTJ} and three-terminal \glspl{FeFET}.
In these cases, the states are read either as resistance or a threshold voltage.
As a result, the read process is non-destructive and better suited for computing systems.

\Gls{MRAM} leverages magnetoresistive effects to store data.
The cell structure is akin to that of \gls{RRAM} and \gls{PCM}, but the difference lies in \gls{MRAM}'s use of two ferromagnetic layers, each capable of holding a magnetic field, separated by a thin tunnel layer.
One of these ferromagnetic layers has a fixed magnetic orientation, while the other layer can alter its magnetic orientation to be either parallel (aligned) or anti-parallel (anti-aligned) relative to the fixed one, representing logical `0' and `1', respectively.
The state is sensed by measuring the resistance, as the two configurations result in different resistances due to the magnetoresistance effects.

Presently, the two primary implementations of \gls{MRAM} are \gls{STT-MRAM} and \gls{SOT-MRAM}.
The cell structures in both cases are similar, with the main difference lying in the methods used to program the cell state.
In \gls{STT-MRAM}, the magnetic orientation of the `free layer' can be modified through spin torque by applying current pulses across the tunnel junction.
On the other hand, in \gls{SOT-MRAM}, the change of state is achieved by applying a current pulse along the heavy metal line (like platinum or tantalum) upon which the tunnel junction is deposited.
The current pulse leads to the accumulation of spin-polarised electrons, resulting in the switching of magnetisation in the free layer.

As previously mentioned, the concept of \gls{ECRAM} has attracted significant attention over the last few years.
\Gls{ECRAM} possesses a transistor-like structure that includes an additional vertical stack comprising a reservoir layer and a solid-state electrolyte.
The channel's conductivity can be modulated by injecting ionised defects into the reservoir layer.
These defects are typically oxygen vacancies or lithium ions, but they could also include other species like protons.
\Gls{ECRAM} offers potential advantages over other \glspl{NVM}, including higher cycling endurance and the ability to control conductance modulation in a linear fashion (when the change in conductance is linearly dependent on the number of applied voltage pulses).
This linear conductance modulation is often important for machine learning applications, especially for training.

Finally, memtransistors, a more recent concept, combine a three-terminal transistor structure with memristor-like capabilities, facilitating changes in the channel's conductance through the application of large source-drain voltages.
These devices typically employ two-dimensional semiconductor materials such as \ce{MoS2} for their channel.
Compared to the other \gls{NVM} technologies mentioned, memtransistors are still in their initial stages of development.

Table~\ref{tab:memory-technologies} provides a comparative analysis of the discussed emerging \gls{NVM} technologies against mature NOR and NAND Flash.

\begin{table}[h]
  \footnotesize
  \centering
  \begin{tabular}{lccccccccc}
    \toprule
    & NOR flash & \thead{NAND\\ flash} & \glsentryshort{RRAM} & \glsentryshort{PCM} & \thead{STT-\\MRAM} & \glsentryshort{FeRAM} & \glsentryshort{FeFET} & \thead{SOT-\\MRAM} & \ce{Li}-ion \\
    \midrule
    On/off ratio & $10^4$ & $10^4$ & $10$--$10^2$ & $10^{2}$--$10^{4}$ & $1.5$--$2$ & $10^{2}$--$10^{3}$ & $5$--$50$ & $1.5$--$2$ & $40$--$10^3$ \\
    \addlinespace[0.5em]
    \makecell[l]{Multilevel\\ operation} & 2 bit & 4 bit & 2 bit & 2 bit & 1 bit & 1 bit & 5 bit & 1 bit & 10 bit \\
    \addlinespace[0.5em]
    \makecell[l]{Write voltage\\(\unit{\volt})} & 10 & 10 & \num{<3} & \num{<3} & \num{<1.5} & \num{<3} & \num{<5} & \num{<1.5} & \num{<1} \\
    \addlinespace[0.5em]
    \makecell[l]{Write time\\(\unit{\nano\second})} & $10^3$--$10^4$ & $10^5$--$10^6$ & \num{<10} & \num{\sim 50} & \num{<10} & \num{\sim 30} & \num{<10} & \num{<10} & \num{<10} \\
    \addlinespace[0.5em]
    \makecell[l]{Read time\\(\unit{\nano\second})} & \num{\sim 50} & \num{\sim e4} & \num{<10} & \num{<10} & \num{<10} & \num{<10} & \num{<10} & \num{<10} & \num{<10} \\
    \addlinespace[0.5em]
    \makecell[l]{Stand-by\\power} & Low & Low & Low & Low & Low & Low & Low & Low & Low \\
    \addlinespace[0.5em]
    \makecell[l]{Write energy\\(\unit{\femto\joule/\bit})} & \num{\sim e5} & \num{10} & \numrange{e2}{e4} & \num{e4} & \num{\sim e2} & \num{\sim e2} & \num{<1} & \num{<e2} & \num{\sim e2} \\
    \addlinespace[0.5em]
    Linearity & Low & Low & Low & Low & None & None & Low & None & High \\
    \addlinespace[0.5em]
    Drift & No & No & Weak & Yes & No & No & No & No & No \\
    \addlinespace[0.5em]
    \makecell[l]{Integration\\density} & High & Very high & High & High & High & Low & High & High & Low \\
    \addlinespace[0.5em]
    Retention & Long & Long & Medium & Long & Medium & Long & Long & Medium & $\cdots$ \\
    \addlinespace[0.5em]
    Endurance & $10^5$ & $10^4$ & $10^5$--$10^8$ & $10^6$--$10^9$ & $10^{15}$ & $10^{10}$ & \num{>e5} & \num{>e15} & \num{>e5} \\
    \addlinespace[0.5em]
    \makecell[l]{Suitability\\for training} & No & No & No & No & No & No & Moderate & No & Yes \\
    \addlinespace[0.5em]
    \makecell[l]{Suitability\\for inference} & Yes & Yes & Moderate & Yes & No & No & Yes & No & Yes \\
    \bottomrule
  \end{tabular}
  \caption{%
    Comparison of different memory technologies for in-memory computing.
    Adapted from Ref.~\cite{mannocci2023inmemorycomputingemerging} under CC~BY~4.0 license.
    }
  \label{tab:memory-technologies}
\end{table}

\subsection{Challenges with analogue hardware}

In contrast to the digital paradigm, where many physical imperfections are masked within a bit representation (`1' or `0'), analogue electronics faces challenges due to the inherent imprecision of non-discrete systems.
Even with minimal amount of nonidealities, it is difficult to encode information with perfect precision, say, number `$4.2$' using a conductance value of \emph{exactly} \qty{4.2}{\milli\siemens}.
But nonidealities do exist, and they can cause \emph{significant} deviations from ideal behaviour.
These include the device becoming stuck in certain conductance states, undergoing changes in conductance over time, showing nonlinear current-voltage characteristics, or displaying nonlinear conductance modulation in response to voltage stimuli, to name a few.
Analogue computing's more fundamental challenge may lie in its reduced precision compared to digital computing, especially when digital systems utilise 16 or more bits representations.

While these issues can be disqualifying for many applications, this might not necessarily be the case for machine learning applications, which often employ reduced precision computing, even within digital systems.
In general, machine learning models have a degree of robustness to small changes, such as noise~\cite{ChSc2017}.
If severe, hardware nonidealities can cause a drop in accuracy, but they do not render these systems \emph{useless}.
Either way, it is important to understand the effects of nonidealities and how they can be mitigated.

\section{Addressing existing challenges}

Novel computer hardware solutions employing analogue devices still suffer from limited precision and unreliability, but both physical and algorithmic techniques can be employed to mitigate these issues.
Unlike digital, the analogue approach involves inherent imprecision.
To add to that, analogue devices, like \gls{RRAM}, often get stuck, experience device-to-device variability or suffer from \IV\ nonlinearity.
Fortunately, fabrication methods and improvements at the circuit level can mitigate some of these nonidealities, while algorithmic techniques can reduce their \emph{effects}.

\subsection{\IV\ nonlinearity}

For linear algebra applications, proportional relationship between voltage and current (i.e. Ohmic behaviour) is preferred.
That is because Ohm's law is used to implement multiplication, as discussed earlier.
However, deviations from this linear relationship do occur, especially in high-resistance devices~\cite{MeMu2017}.

A number of approaches exist at the device and circuit level that help deal or even circumvent the issue of nonlinearity.
During the fabrication of \gls{RRAM} devices, hot-forming step can be adopted, which leads to more linear characteristics~\cite{SuLi2018}.
When programming the devices individually, adopting \gls{1T1R} architecture can help tune memristor conductance precisely, despite any \IV\ nonlinearities~\cite{LiHu2018}.
Alternatively, charge-based accumulation can be adopted, where constant voltage is applied, but the input is encoded into pulse width~\cite{AmAl2020}---this eliminates the dependence on the shape of the \IV\ curve.

\subsection{Faulty devices}\label{sec:faulty-devices}

Some memristive devices may also get stuck in a particular conductance state.
This can happen right after a process like electroforming or even after several successful programming cycles~\cite{JoWa2022}.
In general, the greater the deviation from intended conductance, the greater the potential for negative impact, thus it is crucial to find ways of avoiding---or at least dealing with---faulty devices.

The overall effect of a device getting stuck depends on other devices too, thus this can be used to mitigate the negative effects.
For example, if a device gets stuck, its negative effect may be counteracted by adjusting the conductance of another device in the differential pair~\cite{LiHu2017}.
On occasion, such adjustment happens accidentally by both devices in a differential pair getting stuck.

Alternatively, if faulty devices can be identified before programming, smarter mapping schemes can be employed.
The most important weights can be mapped onto crossbar rows~\cite{GaZh2021} and columns with the lowest incidence of stuck devices.
`Most important' would likely refer to the weights that could have the greatest effect on accuracy.
One approach of identifying such weights is by calculating sensitivity $\Delta w_{i,j}$ for each weight $w_{i, j}$:
\begin{equation}\label{eq:sensitivity}
  \Delta w_{i,j} \defequal - \eta \pd{E}{w_{i,j}}
\end{equation}
where $E$ is the backpropagated loss at the current neuron and $\eta$ is the learning rate.

\subsection{Limited dynamic range}

This nonideality refers to small $G_\mathrm{on}/G_\mathrm{off}$ ratio, which can lead to limited effective precision.
In the context of other nonidealities, like device variability, limited dynamic range can mean that fewer distinguishable states are available.
If each state has certain amount of absolute variability associated with it, then it is clear that larger dynamic range is preferred because it allows to better separate those `fuzzy' states.

The effect of the dynamic range depends \emph{highly} on the application.
If one wishes to use analogue arrays to store digital information, greater dynamic range provides a precision of more equivalent bits.
However, in acceleration of linear algebra operations (and by extension machine learning) such comparisons cannot be drawn so easily.
Because such hardware accelerators rely on analogue computation, the concept of `bits'---although it may be useful---does not apply \emph{directly}.
In analogue contexts, mistakes are any deviations from the intended value, and the magnitude of the mistake is what matters.

For inference applications, a large dynamic range is not crucial.
If a na\"ive mapping scheme is used where the value of a weight ($\in \mathbb{R}$) is represented using a single conductance value, then inaccuracies produced by this imperfect mapping can be addressed with $G_\mathrm{on}/G_\mathrm{off}$ ratio of as low as $3$~\cite{MeJo2019}.
In other contexts, the effect of limited dynamic range cannot be assessed without knowing the nature of other nonidealities, i.e.\ the deviations they cause.

\subsection{Line resistance}\label{sec:line-resistance}

Line resistance is a nonideality stemming from nonzero interconnect resistances in crossbar arrays.
When present, it leads to deviations from the ideal computation of vector-matrix products.
Although the effect on accuracy can be severe, there are both physical and algorithmic techniques for mitigating it.

One of the simplest strategies of reducing the effects of line resistance is increasing the ratio between the resistance of the devices and the resistance of the wires connecting them.
Because resistance is inversely proportional to the cross-sectional area of the wire, one way of decreasing interconnect resistance is by increasing the width of the wires~\cite{LiHa2017}.
Of course, this can be challenging in dense arrays, thus alternative approach is using more conductive materials, e.g.\ \qty{2}{\nano\meter} \ce{Pt} nanofins~\cite{PiLi2019}.
Yet another approach is increasing the resistance of the crossbar devices, but this can sometimes lead to less stable device behaviour~\cite{JoWa2022}.

At the circuit level, multiple techniques that utilise various systematic properties of line resistance can be employed.
For example, a technique called double biasing can make the distribution of electric potentials in the crossbar arrays more symmetric thus reducing the severity of line resistance effects~\cite{HuSt2016}.
Voltage drops in crossbar arrays tend to accumulate as the crossbar size increases, thus splitting them up into smaller ones~\cite{XiGu2016} or even organising them in three-dimensional structures~\cite{XiYa2019} can help.

When considering specific applications that crossbar arrays will be used in, algorithms can be employed to determine optimal mappings from software parameters to physical quantities, like voltage and conductance.
A nonlinear mapping from weights to conductances can be used to counteract the degrading effects of line resistance~\cite{HuSt2016}.
Alternatively, sensitivity analysis (like in Section~\ref{sec:faulty-devices}) can identify the most sensitive weights so they could be mapped closest to the applied voltages, where their contribution would be disturbed the least~\cite{AgLe2019}.
In the specific context of supervised learning, input intensities may be predicted, and the inputs with highest expected intensity (as well as the corresponding weights) can be mapped closest to the outputs in order to minimise the negative effects of line resistance~\cite{JoFr2020,Jo2022}.

\subsection{Programming nonlinearity}

When training networks directly on crossbar arrays, i.e.\ in situ, linear adjustments of conductance are preferred~\cite{BuSh2015}.
To ensure a linear response, the system has to be modified physically.
Some works have proposed adjusting the device structure, typically by introducing additional layers~\cite{WoMo2016,WuWu2018}.
An alternative approach is combining memristive devices with \gls{CMOS} transistors, which help improve the linearity~\cite{AmNa2018}.

\subsection{Random telegraph noise}\label{sec:rtn}

\Gls{RTN} is characterised as unpredictable switching between two or more discrete voltage levels in electronic devices~\cite{PuPa2016} and is often experienced in memristors.
It is more commonly experienced in more resistive devices, thus it can sometimes prevent from using such devices as a way to reduce power consumption or reduce the effects of line resistance, as proposed in Section~\ref{sec:line-resistance}.
To avoid \gls{RTN}, or at least reduce its effects, one typically has to adjust the fabrication process; for example, some works have shown that non-filamentary devices can help reduce this kind of noise~\cite{ChFr2018}.

\subsection{Nonideality-agnostic approaches}

When the specific application where memristive crossbars will be used is known---say, classification using neural networks---a relevant metric---say, accuracy---may be optimised instead of trying to deal with individual nonidealities.
This approach is more technology-agnostic---the nature of nonidealities often varies from one technology to another, but approaches that optimise the metrics relevant to the application tend to be algorithmic and thus more easily transferable.

In machine learning, averaging approaches can make the models more accurate and robust, which is especially relevant if the memristive implementation is prone to nonidealities.
One approach is to use several networks in parallel and average their outputs~\cite{JoFr2020}.
Stability over time can be important too: because some nonidealities (like \gls{RTN}; see Section~\ref{sec:rtn}) are stochastic, averaging over time can reduce their effects~\cite{wanVoltageModeSensingScheme2020}.

Statistical approaches in modern machine learning are based on minimising deviations from ideal behaviour in the training data; this can be extended to incorporate the nonideal effects of the hardware that the model will be implemented on.
In some instances, this can be done by injecting nonideality-agnostic noise during training to make the networks more robust~\cite{huangMethodObtainingHighly2022,yeImprovingRobustnessAnalog2023}.
Alternatively, noise can reflect the nature of the nonidealities, so that the model could more effectively adapt to the various shortcomings of the hardware~\cite{JoWa2022}.

\section{Summary and conclusion}

This article delves into the current state of computing hardware, arguing that the surging computational power needs, driven by machine learning applications, are unlikely to be met with the traditional reliance on transistor scaling.
Data centres, which execute most machine learning tasks, are becoming unsustainably power-hungry.
To add to that, many relatively simple machine learning algorithms cannot even be implemented on low-power edge devices.

We aim to provide a wider perspective on the current state of computer hardware and machine learning.
We briefly outline the fundamentals of machine learning, the currently dominant method for implementing artificial intelligence, the existing hardware and computing paradigms, as well as the inherent limitations present in the von~Neumann architecture and \gls{CMOS} technologies that underpin digital logic.

We then introduce an alternative approach---\glsentrylong{IMC} that is based on analogue computing and emerging \glsentrylong{NVM} devices.
We briefly touch upon emerging memory technologies and provide a comparison focused on computing.
Despite its promising nature, analogue computing is fundamentally susceptible to errors and lacks precision compared to its digital counterparts.
We delve into some of the primary challenges---device and system nonidealities---and explore the main strategies for addressing these issues, from solutions that specifically tackle a particular nonideality to more system-level, algorithmic methods.

Analogue computing is not a new concept---such computers existed before their digital counterparts.
The approach holds significant promise, particularly in terms of energy efficiency in implementing linear algebra, which is integral to today's machine learning algorithms.
Yet, it remains an open question whether such a method will prove practically feasible and what its ultimate scalability potential might be.

If successful, many analogue computing systems might find their use case in accelerating smaller machine learning models, where the cumulative errors may be easier to manage, and the overheads do not render the approach less advantageous compared to digital.
Conversely, it is also plausible that with concurrent hardware-algorithmic development and fundamentally novel algorithms, analogue computing methods may find ways to scale to much larger, more complex models.

Although we have not delved into the subject of brain-inspired or the so-called neuromorphic computing, such methodologies are closely relevant to the discussed topics.
Such approaches, too, aim to enhance the efficiency of artificial intelligence, but they can also potentially \emph{expand} its capabilities.
Several shared themes emerge, like the co-location of memory and compute functions, but neuromorphic computing introduces additional facets, such as sense of locality in time and space, which could be relevant features in next-generation machine intelligence~\cite{mehonic_brains_2023}.

This is an exciting time, where pioneering algorithmic developments are tightly intertwined with advancements in hardware---more so than ever before.
The fundamentals of computing, data processing, and the associated compute hardware are on the cusp of a transformative shift, with the potential for groundbreaking innovations that bring together researchers from a variety of traditional disciplines.

\clearpage

\printbibliography[title=References]

\end{document}